\documentclass{aa}
\usepackage{graphicx}
\def\gta{ \lower .75ex \hbox{$\sim$} \llap{\raise .27ex \hbox{$>$}} }
\def\lta{ \lower .75ex\hbox{$\sim$} \llap{\raise .27ex \hbox{$<$}} }
\begin{document}
%\newcommand{xmm}{XMM-{\it Newton}}

%staubert@hotmail.com,mjlt@star.le.ac.uk

%maraschi@brera.mi.astro.it,henri@obs.ujf-grenoble.fr,matt@fis.uniroma3.it,Suzy.Collin@obspm.fr,dumont@obspm.fr,Martine.Mouchet@obspm.fr,perola@amaldi.fis.uniroma3.it,pferrando@cea.fr,lkoch@cea.fr,Francesco.Haardt@mib.infn.it

\author{P.O.\,Petrucci\inst{1} \and G.\,Henri\inst{1} \and
L.\,Maraschi\inst{2} \and P.\,Ferrando\inst{3} \and G.\,Matt\inst{4} \and
M.\,Mouchet\inst{5} \and C.\,Perola\inst{4} \and S.\,Collin\inst{5} \and
A.M.\,Dumont\inst{5} \and F.\,Haardt\inst{7} \and
L.\,Koch-Miramond\inst{3}}

\institute{Laboratoire d'Astrophysique de Grenoble, BP 43, 38041 Grenoble
  Cedex 9, France \and Osservatorio Astronomico di Brera, Via Brera 28,
  02121 Milano, Italy \and Service d'Astrophysique, DSM/DAPNIA/SAp, CE
  Saclay, 91191 Gif-sur-Yvette Cedex, France \and Dipartimento di Fisica,
  Universit\`a degli Studi ``Roma tre'', via della Vasca Navale 84,
  I-00046 Roma, Italy \and LUTH, Observatoire de Paris, Section de
  Meudon, 92195 Meudon Cedex, France \and Istituto di Astrofisica
  Spaziale, CNR, Via Fosso del Cavaliere 100, I-00133 Roma, Italy \and
  Universit\`a dell'Insubria, Via Lucini 3, 22100 Como, Italy}

\date{Received/ Accepted} \offprints{P.O. Petrucci
\email{petrucci@obs.ujf-grenoble.fr}}

\title{A rapidly variable narrow X-ray iron line in Mkn 841}
\subtitle{}

\abstract{ We report on the detection of a rapidly variable narrow Fe
K$\alpha$ line in Mkn 841. The source has been observed two times by
XMM-Newton and simultaneously with BeppoSAX. The two observations, of
about 10ks long each, were separated by $\sim$ 15 hours. The line flux
reaches a maximum during the first observation and is significantly
reduced in the second one. The continuum shape and flux, instead, keep
roughly constant between the two pointings. Such rapid variability of a
narrow (unresolved by the XMM-pn instrument) line has never been reported
in the past. These results are not easily explained in the standard cold
reflection model where the narrow line component is supposed to be
produced far from the primary X-ray source (e.g. from the torus) and is
thus not expected to vary rapidly. Different interpretations are
discussed.  \keywords{Galaxies: individual: Mkn 841 -- Galaxies: Seyfert
-- X-rays: galaxies}}

\maketitle

%%%%%%%%%%%%%%%%%%%%%%%%%%%%%%%%%%%%%%%%%%%%%%%%%%%%%%%%%%%%%%%
\section{Introduction}
%%%%%%%%%%%%%%%%%%%%%%%%%%%%%%%%%%%%%%%%%%%%%%%%%%%%%%%%%%%%%%%
The fluorescent K$\alpha$ iron line is an important feature in the high
energy spectra of Seyfert galaxies. It is believed to be the result of
fluorescent emission due to photoionization of iron atoms in optically
thick matter illuminated by a compact source of hard X-ray photons. The
study of the iron line properties (ionization state, profile,
variability) indeed provides very useful information on the geometry and
the physical properties of the emitting material (Fabian et
al. \cite{fab00} and references therein).\\ Rapid variability (on less
than a day time scale) of the iron line has been reported in the past for
a few objects (Yaqoob et al. \cite{yaq96}; Vaughan \& Edelson
\cite{vau01}; Wang et al. \cite{wan01}). In all these cases, the line was
relatively broad. This broadening is believed to be a direct probe of
Doppler and gravitational effects that occur in the region very close to
the putative supermassive black hole (hereafter BH). The line variability
would thus be associated with the strong perturbations of the emitting
medium in the vicinity of the BH. Moreover, a narrow iron line appears to
be often, if not always, present in Seyfert 1 galaxies (Pounds \& Reeves
\cite{pou02} and references therein). It is thought to be produced by an
obscuring and remote torus postulated to exist in AGN unification schemes
(Antonucci \cite{ant93}; Ghisellini et al. \cite{ghi94}). This line is
thus not expected to vary on so short time scale.\\ Mkn 841 is a bright
Seyfert 1 galaxy (z=0.0365), one of the rare Seyfert 1 detected by OSSE
at more than 3 $\sigma$ (Johnson \cite{joh97}). It is known for its large
spectral variability (George et al. \cite{geo93}; Nandra et
al. \cite{nan95}), its strong soft excess (this was the first object
where a soft excess was observed, Arnaud et al. \cite{arn85}) and its
variable iron line (at least on a year time scale, George et
al. \cite{geo93}). The latter was observed in some cases with a
relatively large equivalent width (hereafter EW) of about 400 eV (Day et
al. \cite{day90}; Bianchi et al. \cite{bia01}) significantly above the
value predicted by standard cold reflection model (e.g. George \& Fabian
\cite{geo91}).\\ In this paper we present the results from a XMM-Newton
observation of Mkn 841. This source was also observed simultaneously with
the BeppoSAX satellite. We report the presence of a rapidly variable
narrow (i.e. unresolved by the XMM-pn instrument) iron line changing by a
factor of a few in flux on a half day time scale.
%The data suggest also variations on shorter timescales of the order of
%one hour. The peculiarity of these results is that the line has a narrow
%profile, unresolved by the XMM-pn instrument.

%%%%%%%%%%%%%%%%%%%%%%%%%%%%%%%%%%%%%%%%%%%%%%%%%%%%%%%%%%%%%%%
\section{Observation and data analysis}
%%%%%%%%%%%%%%%%%%%%%%%%%%%%%%%%%%%%%%%%%%%%%%%%%%%%%%%%%%%%%%%
Due to operational contingency, the 30ks observation of Mkn 841 with XMM
(Jansen et al. \cite{jan01} and references therein) was split into two
parts, planned the 13th (noted OBS1) and the 14th (noted OBS2) of
February 2001 with $\sim$ 12 and 15 ks exposure time respectively. The
two observations were separated by about 15 hours. \\
In this paper, we will only deal with the EPIC-pn data (we have checked
that the pn and MOS spectra are consistent with each other in the 2-10
keV band). 
%The EPIC-pn and EPIC MOS 2 cameras were operated in Small Window mode
%(hereafter SW), with thin aluminium filters to reject visible light. The
%EPIC MOS 1 was operated in Timing mode. The two RGS and the OM were also
%simultaneously operating. In this paper, we will only deal with the
%EPIC-pn (we have checked that the pn and MOS2 spectra are consistent with
%each other in the 2-10 keV band).
A complete analysis, including all the XMM instruments, is deferred to a
future work. The EPIC-pn camera was operated in Small Window mode
(hereafter SW), with thin aluminium filters to reject visible light. The
EPIC-pn event file was reprocessed from the ODF data files using the {\it
epchain} pipeline task of the XMM Science Analysis System (SASv5.2) and
using the most updated version of the public calibration files. { We note
that SASv5.2 only contained a preliminary model for the pn-chips charge
transfer efficiency. This should not affect our main results presented
here, based on the difference between two spectra, but may change
slightly the spectral shape}.  The source spectra and light curves were
built from photons detected within a 40 arcsec extraction window centered
on the source. X-ray events corresponding to pattern $\le$ 4 were
selected. The background was estimated from an offset position and was
found to be extremely low during the whole observation. The EPIC-pn count
rate was of about 18 cts.s$^{-1}$ for both observations, i.e. well below
the pile-up threshold of $\sim$ 300 cts.s$^{-1}$ in SW mode. { For the
fitting procedure the XMM spectra were binned to have at least 50 counts
per bin.}\\ Mkn 841 was also observed between the 11th and 14th of
February 2001 by the BeppoSAX satellite (Boella et al. \cite{boe97} and
references therein). The source was pointed for a total net time exposure
of $\sim$90, 40 and 20 ks for the MECS, PDS and LECS instruments
respectively. Since we focus our analysis, in this paper, to data above 2
keV, we will not deal with the LECS data any longer.
\begin{figure}
\includegraphics[width=\columnwidth]{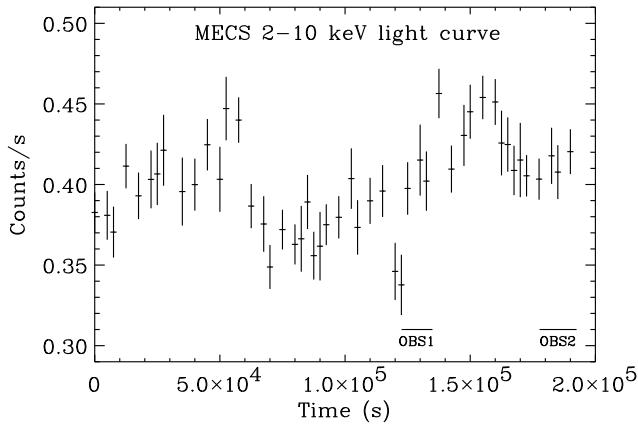}
\caption{MECS light curves (bins of 2500 s). The periods of the
two XMM observations are indicated on the figure.}
\label{fig1}
\end{figure}
The MECS light curve is reported in Fig.~\ref{fig1} with a binning of
2500 s.  Variations of about 20\% are observed.  The periods of the two
simultaneous XMM observations are also reported in this figure. The
EPIC-pn light curves show smooth flux variations of a few percent on time
scale of 1000 s or less.\\
The integrated XMM-Newton flux of Mkn 841 over the 2-10 keV range was
roughly the same for the two EPIC-pn observations at about
1.4$\times$10$^{-11}$ erg.s$^{-1}$.cm$^{-2}$. It slightly differs by
$\sim$10$\%$ (smaller) than the mean 2-10 keV flux measured in the MECS
as it is generally observed between these two instruments (Molendi
private communication).\\
The (2-5 keV)/(5-10 keV) hardness ratio of the MECS being consistent with
no spectral variability during the observation, and in order to maximize
the statistics, we integrated the MECS and PDS spectrum over the entire
observation despite only a part of it was simultaneous with the XMM
one. In the following, all errors refer to 90\% confidence level for 1
interesting parameter ($\Delta\chi^2$=2.7).

%%%%%%%%%%%%%%%%%%%%%%%%%%%%%%%%%%%%%%%%%%%%%%%%%%%%%%%%%%%%%%%
\section{The line variability}
%%%%%%%%%%%%%%%%%%%%%%%%%%%%%%%%%%%%%%%%%%%%%%%%%%%%%%%%%%%%%%%
The XMM-Newton and BeppoSAX instruments need to be carefully normalized
one with each other to correct from calibration uncertainties. The MECS
data were used to perform the cross calibration as follows. The
multiplicative factor between the EPIC-pn and the PDS was obtained by
normalizing the EPIC-pn flux to the MECS spectrum and by applying a
relative normalization factor of { 0.86 (free to vary by $\pm$5\%)}
corresponding to the cross calibration of the MECS and PDS instruments
(Fiore et al. \cite{fio99}). We then obtain a normalization factor of
1.13 and 1.08 for the PDS spectrum relative to the EPIC-pn one for OBS1
and OBS2 respectively.  From now, we will only use the EPIC-pn and PDS
instruments after having checked that the EPIC-pn and MECS spectral shape
were consistent with each other in the 2-10 keV band.\\
\begin{figure}
\begin{tabular}{c}
\includegraphics[angle=-90,width=\columnwidth]{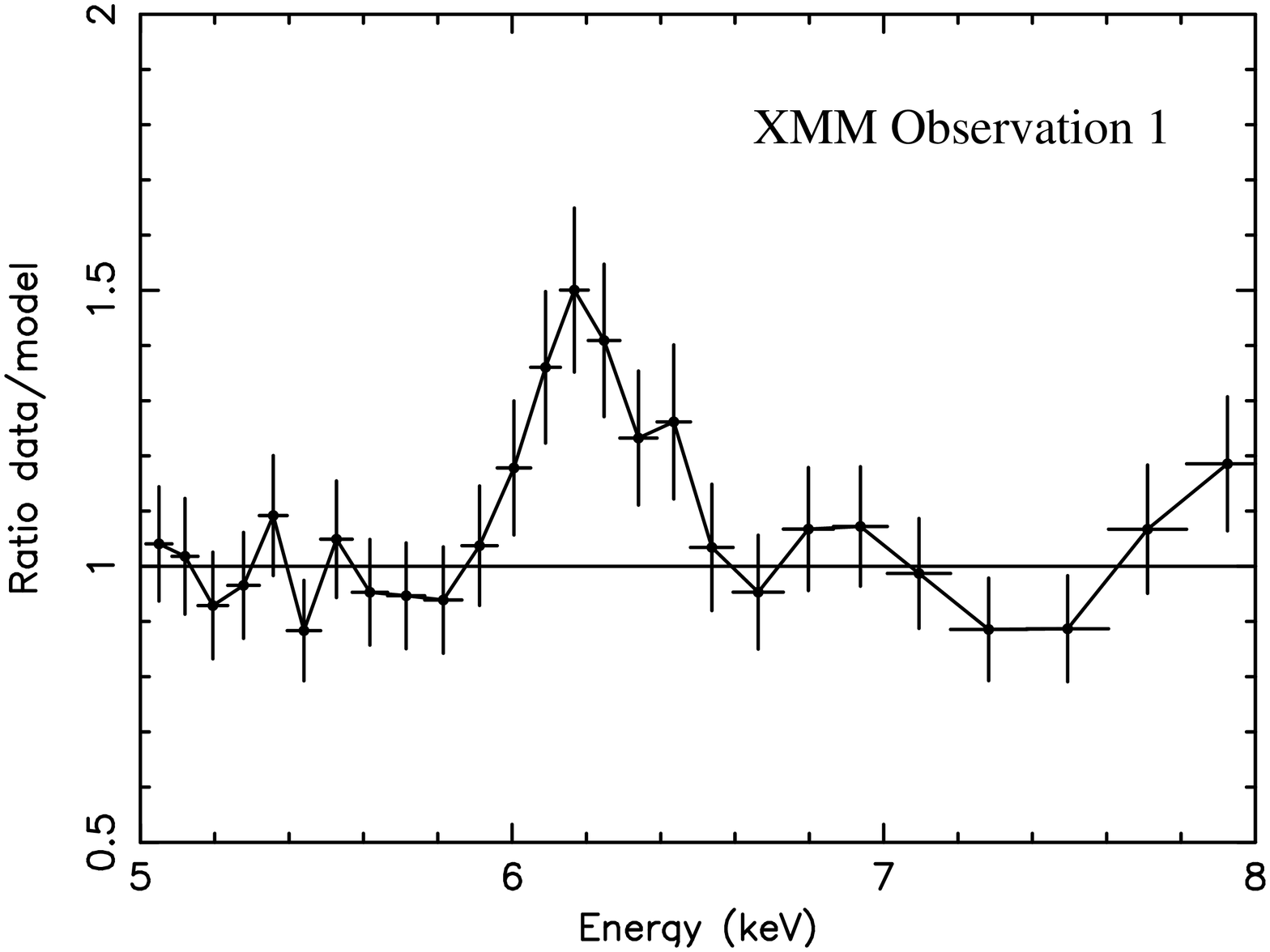}\\
\includegraphics[angle=-90,width=\columnwidth]{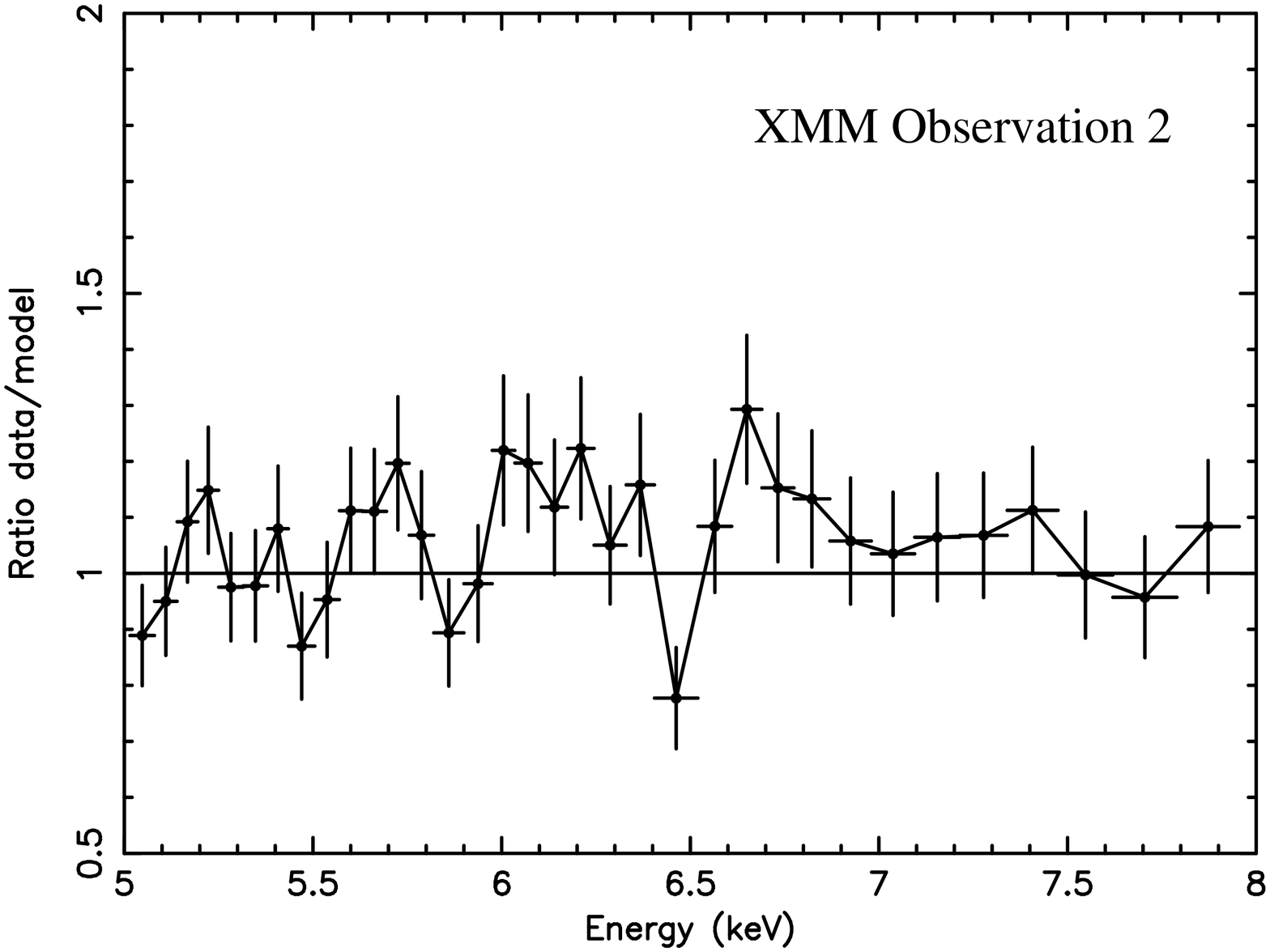}
\end{tabular}
\caption{
Ratios data/model (without line) for OBS1 and OBS2. The {\sc pexrav} fit
was done using the EPIC-pn and PDS data simultaneously. The vertical
dashed line indicates the position of the 6.4 keV neutral iron line in
the source rest frame.}
\label{fig2}
\end{figure}
We have reported in Fig.~\ref{fig2} the ratio between the EPIC-pn spectra
of OBS1 and OBS2 and the best fit {\sc pexrav} model of {\sc xspec}
(Magdziarz \& Zdziarski \cite{mag95}). This best fit was obtained fitting
the EPIC-pn and PDS data simultaneously.  A line is clearly visible near
6.2 keV ($\sim$ 6.4 keV in source rest frame) in OBS1. Indeed the
addition of a Gaussian in the fit of OBS 1 is significant at more than
99\% (following the F-test) with $\Delta\chi^2$=20 for 3 additional
parameters. On the contrary, the addition of a line was not significant
in OBS2 ($\Delta\chi^2$=1).  The best fit parameters of the {\sc
pexrav+gaussian} fits are reported in Table~\ref{tab1}.\\ The energy of
the line detected in OBS1 is consistent with a neutral K$\alpha$ line
$E_{{\mbox{line}}}$=6.41$_{-0.06}^{+0.05}$ keV (source frame), it is
unresolved with a width $\sigma_{\mbox{line}}<$170 eV and has an
equivalent width (EW) of 120$_{-40}^{+50}$ eV (fitting with a {\sc
diskline} model (Fabian et al. \cite{fab89}), the data still constrain
the line to have a narrow profile, the inner disk radius being larger
than $\sim$200 Schwarzschild radii).  We obtain an upper limit of
$\sim$70 eV for the line EW in OBS2 and the best fit line flux is a
factor $\sim$3 weaker than in OBS1.\\
\begin{table}
\begin{tabular}{ccc}
\hline
Observation & OBS1 &OBS2\\
\hline
$\Gamma$ & 2.02$_{-0.04}^{+0.06}$ & 1.99$_{-0.09}^{+0.10}$\\
$R$     & 2.1$_{-0.9}^{+0.4}$   &2.0$_{-1.0}^{+2.7}$\\
$E_c$ (keV)     & $120_{-50}^{+120}$        &$90_{-40}^{+180}$\\
$E_{\mbox{line}}$ (keV)&  6.41$_{-0.06}^{+0.05}$& 6.4$^*$\\
$\sigma_{\mbox{line}}$ (keV)&$<$0.17& 0.1 (fixed)\\
$F_{\mbox{line}}$  (10$^{-6} $ph.cm$^{-2}$.s$^{-1}$)&18$_{-6}^{+7}$ & $<$11\\
EW (eV) & 120$_{-40}^{+50}$ & $<$ 70 \\
$\chi^{2}/dof$ &208/243&328/312\\
\hline
\end{tabular}
\mbox{\small{$^*$ forced to be in the 6-7 keV energy range}}
\caption{Best fit parameters for OBS1 and OBS2 when fitting with the {\sc
pexrav+gaussian} model above 2 keV. The line energy is in the source rest
frame.}
\label{tab1}
\end{table}
\begin{figure}
\includegraphics[width=\columnwidth]{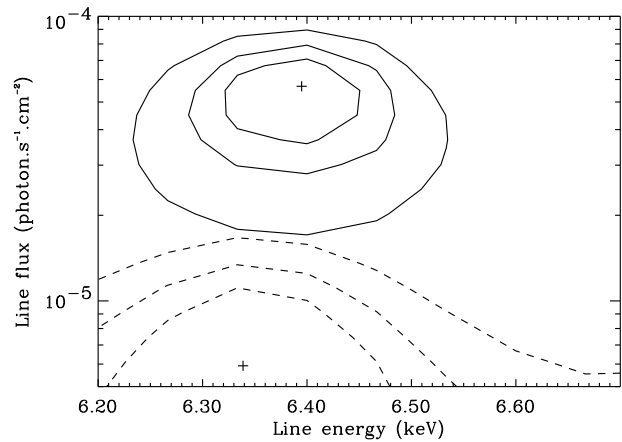}
\caption{Contour plot (68\%, 90\% and 99\% confidence level) of the line
flux vs. the line energy during the line flux maximum of OBS1 (solid
line) and the total OBS2 (dashed line). Cf. text for details.}
\label{fig3}
\end{figure}
We then checked for line flux variations on smaller time scales. For that
purpose, we studied the light curves of the flux ratios (6-6.6 keV)/(3-5
keV) and (6-6.6 keV)/(7-10 keV). These ratios are expected to be
sensitive to the line flux variations with respect to the continuum shape
in each side of the line. The line interval is smaller than the others to
be sensible enough to the line flux variability. The light curves show a
significant increase of both flux ratios during OBS1 where the line flux
reaches a maximum { of $\sim$5$\times$10$^{-5}$
ph.cm$^{-2}$.s$^{-1}$}. Fits of the different EPIC-pn spectra along the
light curves (using a temporal bining of 1800 s) show that the best fit
line flux varies by a factor $\sim$ 6 during the total observation (the
lowest flux being in OBS2). The line width is always consistent with zero
i.e. the line is always unresolved by the instrument.\\ We have plotted
on Fig.~\ref{fig3} the contour plots of the line flux versus the line
energy (with contours at 68, 90 and 99\%) obtained during the
sub-interval of OBS1 where the line flux reaches its maximum (solid
contours) and for the total OBS2 (dashed contours). These contour plots
have been obtained using the EPIC-pn data alone fitted with {\sc pexrav +
gaussian}. The line width was kept fixed to the value of 0.1 keV and we
also fixed the reflection normalization parameter $R$ to its best fit
value. From these contours, the hypothesis of the constancy of the line
can be rejected at more than 99.9\% { (it is rejected at 99.5\% if we use
the total OBS1)}. {Unfortunately, the BeppoSAX/PDS data does not provide
enough counts to test whether the reflection also varied significantly
during the pointing.}

%%%%%%%%%%%%%%%%%%%%%%%%%%%%%%%%%%%%%%%%%%%%%%%%%%%%%%%%%%%%%%%
\section{Discussion}
%%%%%%%%%%%%%%%%%%%%%%%%%%%%%%%%%%%%%%%%%%%%%%%%%%%%%%%%%%%%%%%
Two observational facts of the present data are not easy to account for.
Firstly the {\it narrow} line varies rapidly by a factor of a few while
the continuum changes by only $\sim$ 10-20\%. Secondly the reflection
seems relatively large, $R>1$, while the line EW is relatively small
$<170$ eV ({it is worth noting that fitting the BeppoSAX data above 2 keV
gives also a large $R$=2.5$^{+2.5}_{-1.3}$ and a small $EW<$ 100
eV}). These points are in contradiction with the standard cold reflection
model. In this model, the line width is due to the motion of the emitting
matter in the BH gravitational potential. In this case, a fluorescent
neutral iron line with a width of about 150 eV { ($\sim$ 14000
km.s$^{-1}$)}, like in our case, is expected to be emitted at a distance
${r\simeq 10^{17}M_8/q}$ cm where the BH mass $M_{\mbox{bh}}=10^8M_8$
solar masses and $q$ depends on the distribution of the orbital shape and
inclination of the matter spiraling around the BH ($q$ is expected to be
larger than 1, Krolik 2001).\\ On the other hand, a variability time
scale of $\sim$ 10 hours constrains the size $d$ of the line emitting
region to about 10 light hours, i.e. ${d\simeq 10^{15}}$cm. The solid
angle sustained by the emitting region, as seen by the X-ray source, is
then $\Delta\Omega\simeq\pi d^2/r^2\simeq3\times 10^{-4}q^2M_8^{-2}$. The
observed values of $R=\Delta\Omega/2\pi>$1 then imply
$qM_8^{-1}\gta140$. However, the fact that we do not observe any flare in
the MECS light curve, which would be at the origin of the line
variability, constrains $r$ to be larger than about 1 light day, meaning
$qM_8^{-1}\lta$40 in contradiction with the previous estimation and
independently of the BH mass. Modifications of this standard view are
thus clearly needed.\\ It is first possible that the observed line
variability is not controlled by the continuum itself but by an intrinsic
variability of the reflecting material due to, e.g. thermal
instabilities. These instabilities may change the ionization state of the
upper layers of the matter. The strength of the neutral iron line,
expected to be produced in the deeper layers, could then be degraded by
the skin scattering properties (Nayakshin \& Kallman \cite{naykal01};
Petrucci et al. \cite{pet01}). On the other hand, Nayakshin \& Kazanas
(\cite{naykaz02}) also showed that, in response to changes in the X-ray
flux, the time for re-adjustment of the hydrostatic balance of the
illuminated gas may be longer than the light-travel time between the
primary X-ray source and the reflecting material. The line variability
observed here may then result from an X-ray flare produced { in the inner
disc region {\it but}} occuring some time before the start of the XMM and
even BeppoSAX observations. We note however that the change of ionization
state of the emitting region suggested above would probably produce a
ionized iron line which is not detected.\\ {Alternatively, the line
emission may have been affected by the continuum outside of the observed
band pass. Variations of the high--energy cut--off and/or of the UV--EUV
emission can indeed alter the ionization structure of the disc surface
without obvious changes in the 2--10 keV continuum (Vaughan \& Edelson
\cite{vau01})}. The EPIC data down to $\sim$0.5 keV do not show
significant variation between OBS1 and OBS2 but a more precise study of
the RGS and OM data will help to better test this hypothesis.\\ Then, if
we admit that the line variability can exist without significant
variations of the continuum in the XMM energy range, the distance $r$ of
the emitting line region estimated above could be compatible with a half
day variability time scale (or less) if the BH mass of Mkn~841 is
relatively small, $< 10^6$ solar masses. It is however significantly
smaller than the $\sim 10^8$ solar masses recently estimated by Laor
(2001).\\ Interestingly, a narrow profile is also expected, independently
of the BH mass, if the flare at the origin of the line variability occurs
in the inner regions but close to the disc and thus illuminates a small
fraction of its surface (Nayakshin \& Kazanas \cite{naykaz01}; Yaqoob
\cite{yaq01}). However, in this case the line is generally expected to be
shifted due to relativistic effects, which is not observed.\\ A less
``standard'' explanation could be the crossing of an obscuring thick
cloud in between the X-ray source and the reflecting matter. This cloud
needs to be close enough to the X-ray source to rapidly eclipse a large
part of its emission. It can explain why the continuum does not change
between the two XMM observations since the cloud may not cross the line
of sight. We then would expect the reflection hump to vary, an hypothesis
which is unfortunately not testable with the present data.

%%%%%%%%%%%%%%%%%%%%%%%%%%%%%%%%%%%%%%%%%%%%%%%%%%%%%%%%%%%%%%%%
\section{Conclusion}
%%%%%%%%%%%%%%%%%%%%%%%%%%%%%%%%%%%%%%%%%%%%%%%%%%%%%%%%%%%%%%%%
The neutral narrow iron line of Mkn 841 appears to be highly variable on
a timescale of 10 hours or less. This cannot be easily explained in the
standard cold reflection model framework. None of the explanations
proposed here appear completely conclusive. A consistent interpretation
of these results has to concile the presence of this variable narrow and
apparently neutral line, a constant (in flux and shape) underlying X-ray
continuum and a large ($>$1) reflection normalization. New data are
clearly needed to confirm this observation and to understand the origin
of the line variability.

\begin{acknowledgements}
We thanks the referee, S. Vaughan, for his useful comments.
\end{acknowledgements}

%\bibliographystyle{aa}
%\bibliography{/home/pop/PAPIERS/ref.bib}

\end{document}